\documentstyle[IJFCS,amstext,latexsym,amssymb,emp,cite]{article}

\newcommand{\C}[0]{{\mathbb{C}}}
\newcommand{\F}[0]{{\mathbb{F}}}
\newcommand{\N}[0]{{\mathbb{N}}}
\newcommand{\R}[0]{{\mathbb{R}}}
\newcommand{\Z}[0]{{\mathbb{Z}}}

\newcommand{\CNOT}[0]{{\rm CNOT}}

\def\ket#1{\left|#1\right>}

\def\qed{\hfill $\Box$}

\newtheorem{keylemma}[theorem]{Key Lemma}
\newtheorem{example}[theorem]{Example}

\newcommand{\diag}[0]{{\rm diag}}
\newcommand{\cas}[0]{{\rm cas}}
\newcommand{\zirc}[0]{{\rm circ}}
\newcommand{\onemat}[0]{\mbox{\bf 1}}
\newcommand{\zeromat}[0]{\mbox{\bf 0}}
\newcommand{\DFT}[0]{F}

\newcommand{\nix}[1]{}

\begin{document}
%% print out the publisher copyright heading
\copyrightheading

%% use symbolic footnote
\symbolfootnote

%% use normal text like skip (13pt)
\textlineskip

\begin{center}

%% print out titles in IJFCS format
\fcstitle{QUANTUM SOFTWARE REUSABILITY}

\vspace{24pt}

{\authorfont ANDREAS KLAPPENECKER}

\vspace{2pt}

%% use smaller line skip here
\smalllineskip
{\addressfont Department of Computer Science,\\ Texas A\&M University,\\
College Station, TX 77843-3112, USA}

\vspace{10pt}
and

\vspace{10pt}
{\authorfont MARTIN R\"OTTELER}

\vspace{2pt}
\smalllineskip
{\addressfont Department of Combinatorics and Optimization\\
University of Waterloo, Waterloo, Ontario, Canada N2L 3G1}

\vspace{20pt}
%% authors need not care about this
\publisher{(\today)}{(revised date)}{Jozef Gruska}

\end{center}

\alphfootnote

%% abstract environment
\begin{abstract}
The design of efficient quantum circuits is an important issue in
quantum computing. It is in general a formidable task to find a highly
optimized quantum circuit for a given unitary matrix.  We propose a
quantum circuit design method that has the following unique
feature: It allows to construct efficient quantum circuits in a
systematic way by reusing and combining a set of highly optimized
quantum circuits. Specifically, the method realizes a quantum circuit
for a given unitary matrix by implementing a linear combination of
representing matrices of a group, which have known fast quantum
circuits.  We motivate and illustrate this method by deriving
extremely efficient quantum circuits for the discrete Hartley
transform and for the fractional Fourier transforms.  The sound
mathematical basis of this design method allows to give meaningful and
natural interpretations of the resulting circuits.  We demonstrate
this aspect by giving a natural interpretation of
known teleportation circuits.

\end{abstract}

%%%%%%%%%%%%%%%%%%%%%%%%%%%%%%%%%%%%%%%%%%%%%%%%%%%%%%%%%%%%
%
% Section: Introduction
%
%%%%%%%%%%%%%%%%%%%%%%%%%%%%%%%%%%%%%%%%%%%%%%%%%%%%%%%%%%%%

\textlineskip
\empprelude{input qcg; prologues := 0;}
\begin{empfile}
\section{Introduction}
The worldwide efforts to build a viable quantum computer have
one source of motivation in common: The potential to solve certain
problems faster on a quantum computer than on any classical computer.
There are a number of ways to specify quantum algorithms but the
formulation of quantum algorithms as a uniform family of quantum
circuits is the most popular choice. 

Deriving an efficient quantum circuit for a given unitary matrix is a
daunting, and, frustratingly, often impossible, task.
There exist a small number of efficient quantum circuits, and
even fewer quantum circuit design methods. If efficient quantum
circuits are rare and difficult to derive, then is only natural to try
to reuse these quantum circuits in the construction of other quantum
circuits. We present in this paper a new design principle for quantum
circuits that is exactly based on this idea.

Suppose that we want to realize a given unitary matrix $A$ as a
quantum circuit. Suppose that we know a number of quantum circuits
realizing unitary matrices $D(g)$ of the same size as $A$.  We choose
a small subset of these unitary matrices such that the algebra
generated by the matrices $D(g)$ contains $A$.  Roughly speaking, if
the generated algebra has some structure, e.g. is a finite dimensional
(twisted) group algebra, then we are able to write down a quantum
circuit realizing~$A$, which reuses the implementations of the
matrices $D(g)$.

As a motivating example serves the discrete Hartley transformation,
which is a variant of the discrete Fourier transform defined over the
real numbers. In Section~\ref{motivation}, we show how the Hartley
transforms can be realized by combining quantum circuits for the
discrete Fourier transform and its inverse.  We generalize the idea
behind this construction in the subsequent sections.

An essential ingredient of our method are circulant matrices and
certain block-diagonal matrices, which are introduced in
Sections~\ref{circulant} and~\ref{group-indexed}.  In
Section~\ref{design} we present the main result of this paper.  We
show how to derive a quantum circuit for a unitary matrix $A$, which
can be expressed as a linear combination $U = \alpha_1 D(g_1) + \ldots
+ \alpha_n D(g_n)$ of unitary matrices $D(g_i)$ with known quantum
circuits. For ease of exposition, we do not state the theorem in full
generality; the generalizations of the method are discussed in
Sections~\ref{sec:generalization} and~\ref{projCirculants}.

There is a relation between Kitaev's method for eigenvalue estimation
of unitary operations and the present method which is explored in
Section \ref{kitaevRelation}. Section \ref{examples} demonstrate the
design principle with the help of some simple examples. We revisit the
Hartley transform in Section~\ref{hartley}, and discuss fractional
Fourier transforms in Section~\ref{fractional}. We give an
interpretation of the well-known teleportation circuit in terms of
projective circulants in Section~\ref{teleport}.

\normalsize

\medskip
\textit{Notations.} We denote by $\Z$, $\Z/n\Z$, $\R$, and $\C$ the ring
of integers, the ring of integers modulo $n$, the field of real
numbers, and the field of complex numbers, respectively.  The group of
unitary $n\times n$ matrices is denoted by ${\cal U}(n)$. We denote
the identity matrix in ${\cal U}(n)$ by $\onemat_n$.

%%%%%%%%%%%%%%%%%%%%%%%%%%%%%%%%%%%%%%%%%%%%%%%%%%%%%%%%%%%%
%
% Section: Background
%
%%%%%%%%%%%%%%%%%%%%%%%%%%%%%%%%%%%%%%%%%%%%%%%%%%%%%%%%%%%%

\section{Background}
\label{basics}
We consider quantum computations that manipulate the state of~$n$
two-level systems. A two-level system has two clearly distinguishable
states $\ket{0}$ and $\ket{1}$, which are used to represent a bit. We
refer to such a two-level system as a quantum bit, or shortly a qubit.
The state of $n$ quantum bits is mathematically represented by a
vector in $\C^{2^n}$ of norm~1.  We choose a distinguished orthonormal
basis of $\C^{2^n}$ and denote its basis vectors by
$\ket{x_{n-1},\dots,x_0}$, where $x_i\in \{0,1\}$ with $0\le i<n$.

A quantum gate on $n$ qubits is an element of the group of unitary
matrices~${\cal U}(2^n)$. We will use single-qubit gates and
controlled-not gates. A {\em single-qubit gate} acting on qubit $i$ is
given by a matrix of the form $U^{(i)} = \onemat_{2^{n-i-1}} \otimes U
\otimes \onemat_{2^{i}}, $ with $U\in {\cal U}(2)$.
A {\em controlled-not gate} with control qubit $i$ and target 
qubit $j$ is defined by 
\[
\ket{x_{n-1}, \dots, x_{j+1}, x_j, x_{j-1}, \dots, x_0}
\mapsto
\ket{x_{n-1}, \dots, x_{j+1}, x_i\oplus x_j, x_{j-1}, \dots, x_0},
\]
where $\oplus$ denotes addition modulo 2. We denote this gate by ${\rm
CNOT}^{(i,j)}$. We will refer to single-qubit gates and controlled-not
gates as {\em elementary gates}.

It is well-known \cite{BBC+:95} that the single-qubit gates and the 
controlled-not gates are universal, meaning the set 
$${\cal G} = \{ U^{(i)}, {\rm CNOT}^{(i, j)}\mid U\in{\cal U}(2),\
i,j\in \{1,\dots, n\},\ i\neq j \}$$ generates the unitary group
${\cal U}(2^n)$. In other words, each matrix $U \in {\cal U}(2^n)$ can
be expressed in the form $U = w_1 w_2 \ldots w_k$ with $w_i\in {\cal
G}$, $0\le i<k$.  Of special interest are the shortest possible words
for $U$. We denote by $\kappa(U)$ the smallest $k$ such that there
exists a word $w_1w_2\cdots w_k$, with $w_i\in {\cal G}$, $0\le i<k$,
such that $U = w_1 w_2 \ldots w_k$.

The complexity measure $\kappa$ turns out to be rather rigid. It is
desirable to allow a variation which gives additional freedom.  We say
that a unitary matrix $V$ realizes $U$ with the help of ancillae
provided that $V$ maps $\ket{0}\otimes \ket{x} \mapsto \ket{0}\otimes
U\ket{x}$ for all $\ket{x} \in \C^{2^n}$.  We define $\kappa_{{\rm
anc}}(U)$ to be the minimum $\kappa(V)$ of all unitary matrices $V$
realizing $U$ with the help of ancillae.

As examples, we mention the following bounds on the complexity for well-known 
transforms acting on $n$ quantum bits: the
Hadamard transform $\kappa(H_2^{\otimes n}) = O(n)$; 
the discrete Fourier transform
$\kappa(\DFT_{2^n}) = O(n^2)$ when realized without ancillae
\cite{ABH+:2001,Shor:94}, and $\kappa_{{\rm anc}}(\DFT_{2^n}) = O(n
(\log n)^2 \log \log n)$ when realized with ancillae~\cite{CW:2000}.
Various unitary signal transformations with fast quantum
realizations can be found
in~\cite{Ettinger:00,Hoyer:97,Klappenecker:99,KR:2001,PRB:99}.

%%%%%%%%%%%%%%%%%%%%%%%%%%%%%%%%%%%%%%%%%%%%%%%%%%%%%%%%%%%%
%
% Section: Motivation
%
%%%%%%%%%%%%%%%%%%%%%%%%%%%%%%%%%%%%%%%%%%%%%%%%%%%%%%%%%%%%

\section{A Motivating Example}
\label{motivation}
Assume that we have already found an efficient quantum circuit for a
given unitary matrix $U\in {\cal U}(2^n)$ with $O(n^c)$ quantum gates,
$c$ some constant.  We would like to find an efficient quantum circuit
for a polynomial function $f(U)$ of $U$, allowing ancillae qubits. 
If we succeed, then this would prove that 
$\kappa_{\rm anc}(f(U)) \in O(n^{k})$
for some constant $k\ge 0$.

As an example, consider the {discrete Hartley transform} $A_N\in {\cal U}(N)$ 
of length $N\in \N$, which is defined by
\[ A_N := \frac{1}{\sqrt{N}}\Big[ \cas\Big(\frac{2\pi k l}{N}
\Big) \Big]_{k,l=0,\ldots, N-1},
\]
where the function $\cas: \R \rightarrow \R$ is defined by
$\cas(x) := \cos(x) + \sin(x)$. 
The discrete Hartley transform is well-known in classical
signal processing, cf.~\cite{beth:89,Bracewell:79}.
If we denote the discrete Fourier transform by $F_N$, then
\[ A_N = \left(\frac{1-i}{2}\right) \; F_N 
    + \left(\frac{1+i}{2}\right) \; F_N^3
\] 
is an immediate consequence of the definitions.

Let $N=2^n$. We will
now derive an efficient quantum circuit implementing the Hartley
transform~$A_N$ with one auxiliary quantum bit.

% Schaltkreis zur Hartleytransformation
% -------------------------------------
\nix{\begin{figure}[hbt]
\input{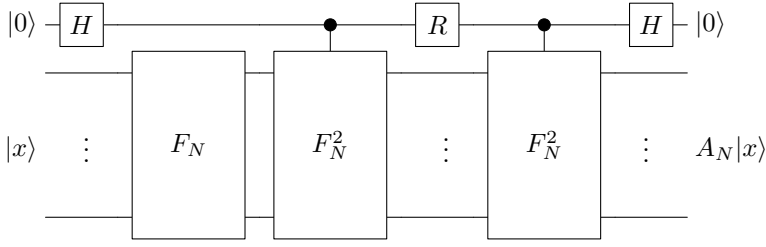}
\bigskip
\caption{
Circuit realizing a quantum Hartley transform}
\end{figure}
}
\begin{figure}[hbt]
\begin{center}
\begin{emp}(50,50)
  setunit(1.6mm);
  qubits(5);
  dropwire(1,2);
  label.lft(btex $|0\rangle$ etex, (QCxcoord,QCycoord[2]));
  label.lft(btex $|x\rangle$ etex, (QCxcoord,QCycoord[0]+0.9cm));
  label.lft(btex $\vdots$ etex, (QCxcoord+0.7cm,QCycoord[0]+1.0cm));

  gate(gpos 2, btex $H$ etex);
  circuit(1.5cm)(gpos 0,1,btex $F_N$ etex);
  circuit(1.5cm)(icnd 2, gpos 0,1,btex $F_N^2$ etex);
  label(btex $\vdots$ etex, (QCxcoord+0.6cm,QCycoord[0]+1.0cm));
  gate(gpos 2, btex $R$ etex);
  circuit(1.5cm)(icnd 2, gpos 0,1,btex $F_N^2$ etex);
  gate(gpos 2, btex $H$ etex);
  label.rt(btex $\vdots$ etex, (QCxcoord-0.7cm,QCycoord[0]+1.0cm));
  label.rt(btex $A_N|x\rangle$ etex, (QCxcoord,QCycoord[0]+0.9cm));
  label.rt(btex $|0\rangle$ etex, (QCxcoord,QCycoord[2]));
\end{emp}
\end{center}
\caption{\label{hartleyCirc} 
Circuit realizing a quantum Hartley transform}
\end{figure}

\begin{lemma}\label{factorHart}
The discrete Hartley transform can be realized by the circuit shown in 
Figure \ref{hartleyCirc}, where $R$ denotes the unitary circulant matrix
\[ 
R := \frac{1}{2}
\left( \begin{array}{rr} 1-i & 1+i \\ 1+i & 1-i \end{array} \right)
\]
and $H$ the Hadamard transform.
\end{lemma}
{\bf Proof.} Let $\widehat{F}_N=\Lambda_1(F_N)$ denote the unitary matrix
effecting a discrete Fourier transform on the $n$ least
significant bits if the most significant (ancilla) bit is set; 
in terms of matrices $\hat{F}_N=\onemat_N \oplus F_N$. Similarly, 
let $\widehat{F}_N^2=\onemat_N \oplus F_N^2$.

We now show that the circuit shown in Figure~\ref{hartleyCirc}
computes the linear transformation $\ket{0}\ket{x} \mapsto \ket{0} A_N
\ket{x}$ for all vectors $\ket{x}\in \C^n$ of unit length.  Proceeding
from left to right in the circuit, we obtain
\begin{eqnarray*}
\ket{0}\ket{x} 
& \stackrel{H}{\longmapsto} & \frac{1}{\sqrt{2}} (\ket{0} + \ket{1})
\ket{x} \\
& \stackrel{F_N}{\longmapsto} & \frac{1}{\sqrt{2}} (\ket{0} + \ket{1})
F_N \ket{x} \\
& \stackrel{\widehat{F}_N^2}{\longmapsto} & 
\frac{1}{\sqrt{2}} \ket{0} F_N \ket{x} + 
\frac{1}{\sqrt{2}} \ket{1} F_N^3 \ket{x} \\
& \stackrel{R}{\longmapsto} & 
\frac{1}{\sqrt{2}}\ket{0} \left(\frac{1}{2} (1-i) F_N + 
\frac{1}{2} (1+i) F_N^3\right) \ket{x}  \\
& \phantom{\stackrel{R}{\longmapsto}} &
+\frac{1}{\sqrt{2}}\ket{1} \left(\frac{1}{2} (1+i) F_N + \frac{1}{2} (1-i)
F_N^3 \right) \ket{x}  \\
& = & \frac{1}{\sqrt{2}} \ket{0} A_N \ket{x} +
\frac{1}{\sqrt{2}} \ket{1} F_N^{-2} A_N \ket{x}\\
& \stackrel{\widehat{F}_N^2}{\longmapsto} & 
\frac{1}{\sqrt{2}} (\ket{0} + \ket{1}) A_N \ket{x} \\
& \stackrel{H}{\longmapsto} & \ket{0} A_N \ket{x}
\end{eqnarray*}
as desired. Note that we have used the property that the discrete
Fourier transform has order four, i.\,e., $F_N^4 = \onemat$. \qed
\smallskip

We cast this factorization and the corresponding complexity cost in
terms of elementary quantum gates in the following theorem.

\begin{theorem} The discrete Hartley transform 
$A_{2^n}\in {\cal U}(2^n)$ can be implemented with 
$O(n (\log n)^2 \log \log n)$ quantum gates on a quantum computer.
\end{theorem}
\smallskip
\noindent
{\bf Proof.} Recall that the discrete Fourier transform $F_{2^n}$ can
be implemented with $O(n (\log n)^2 \log \log n)$ quantum gates, 
see~\cite{CW:2000}. The claim is an immediate consequence of 
Lemma~\ref{factorHart}.~\qed
\medskip

We pause here to discuss some noteworthy features of the preceding
example. We notice that the discrete Fourier transform $F_{2^n}$
satisfies the relation $F_{2^n}^4=1$, and that all powers
$\onemat_{2^n}, F_{2^n}, F_{2^n}^2, F_{2^n}^3$ have fast
implementations with known quantum circuits.  We constructed the
Hartley transform as a linear combination of some of those powers,
namely as a linear combination of $F_{2^n}$ and $F_{2^n}^3$ using the
circuit shown in Figure~\ref{hartleyCirc}. A nice feature of this
circuit is that any improvement in the design of quantum algorithms
for the discrete Fourier transform will directly lead to an improved
performance of the discrete Hartley transform, since the circuits for
the discrete Fourier transform a simply {\em reused}\/ in the Hartley
transform circuit.

We will generalize this idea in the following sections.  The methods
are much more general. We will even be able to combine several
different circuits, assuming that some regularity conditions are
satisfied. The factorization of the Hartley transform implied by Lemma
\ref{factorHart} will be obtained as a special case of this more
general theory in Section \ref{hartley}.

%%%%%%%%%%%%%%%%%%%%%%%%%%%%%%%%%%%%%%%%%%%%%%%%%%%%%%%%%%%%
%
% Section: Circulant Matrices
%
%%%%%%%%%%%%%%%%%%%%%%%%%%%%%%%%%%%%%%%%%%%%%%%%%%%%%%%%%%%%

\section{Circulant Matrices}\label{circulant}
Our goal is to derive a circuit implementing linear combinations of
matrices. This is not an easy task, because all operations need to be
unitary. We assume that the algebra generated by the matrices has some
structure which we can exploit when deriving the circuit.  Our
approach will be particularly successful when the generated algebra
$\mathcal{A}$ is a finite dimensional (twisted) group algebra.  In
this case, we can write down a {\em single}\/ generic circuit which is
able to implement {\em any}\/ unitary matrix $U$ contained in
$\mathcal{A}$.  In the case of a group algebra, a group-circulant
determines which matrix $U$ is implemented by the generic
circuit. 

Recall the definition of a group-circulant~\cite{CB:93,Davis:79}:

\begin{definition}
Let\/ $G$ be a finite group of order $d$. Choose an ordering
$(g_1,\ldots, g_{d})$ of the elements of $G$ and identify the standard
basis of $\C^d$ with the group elements of $G$. Let
$\ket{c} = \sum_{g \in G} c_g \ket{g} \in {\C}^d$ denote a vector indexed
by the elements of $G$.  Then the $d \times d$-matrix
\begin{equation}
% \zirc_G(\ket{c}) := \sum_{h \in G} \sum_{g \in G}
% c_g \ket{h}\bra{h^{-1} g}
\zirc_G(\ket{c}) := (c_{g^{-1}h})_{g,h\in G}
\end{equation}
is a group-circulant for the group $G$. 
\end{definition}

The following example covers the important special case of cyclic
circulants.

\begin{example}
Let $G={\Z}_d$ be the cyclic group of order $d$ generated by $x$
and the elements of $G$ ordered according to $(1, x, x^2,
\ldots, x^{d-1})$. The circulant corresponding to $\ket{c} :=
\sum_{i=0}^{d-1} c_i \ket{x^i} \in {\C}^d$ takes the form
\[
\left(\begin{array}{cccc}
 c_0&\cdots&c_{d-2}&c_{d-1}\\
 c_{d-1}&\cdots&c_{d-3}&c_{d-2}\\
  \vdots  &\ddots&\ddots& \vdots\\
 c_1&\cdots&c_{d-1}&c_0
\end{array}\right).
\]
We see that each row is obtained from the previous one by a cyclic
shift to the right.
\end{example}

The following crucial observation connects group-circulants with the
coefficients $\alpha_g$ in linear combinations. The linear
independence of the representing matrices $D(g)$ of a finite group $G$
ensures that the circulant $(\alpha_{g^{-1}h})_{g,h\in G}$ is a unitary
matrix. 

\begin{keylemma}
Let $n$ be a positive integer. 
Let\/ $\{D(g) : g \in G\}$ be a set of linearly independent unitary
matrices which form a finite subgroup of ${\cal U}(n)$. 
Furthermore, let $A\in {\cal U}(n)$ be a linear combination of the matrices $D(g)$, 
\[ A = \sum_{g \in G} \alpha_g D(g),\]
with certain coefficients $\alpha_g \in \C$. Then the associated group
circulant matrix $C_A := (\alpha_{g^{-1}h})_{g,h\in G}$ is unitary.
\end{keylemma}

\noindent
{\bf Proof.} Multiplying $A$ with $A^\dagger$ yields
\begin{eqnarray*}
A \cdot A^\dagger & = & \Big(\sum_{g \in G} \alpha_g D(g) \Big) \cdot
\Big(\sum_{h \in G} \overline{\alpha_h} D(h)^\dagger \Big) \\
& = & \sum_{g \in G} \sum_{h \in G} \alpha_{g} 
\overline{\alpha_h} D(g) D(h)^\dagger \\
& = & \sum_{g \in G} \Big(\sum_{h \in G} \alpha_{gh} 
\overline{\alpha_h}\Big) D(g).
\end{eqnarray*}
Since the matrices $D(g)$ are linearly independent, it is possible to compare
coefficients with $A \cdot A^\dagger = \onemat_n$, which shows that
\[ \sum_{h \in G} \alpha_{gh} 
\overline{\alpha_h} = \delta_{g, e}
\]
holds, where $\delta_{i, j}$ denotes the Kronecker-delta. In other
words, the rows of the circulant matrix $C_A$ are orthogonal.  \qed
\medskip

\textit{Remark.} If the representing matrices $D(g)$ are not linearly
independent, then the group circulant is in general not unitary.  In
fact, it is not difficult to see that for {\em each}\/ unitary matrix
$A$ there {\em is}\/ a choice of coefficients $\alpha_g$ such that the
group-circulant is not unitary. However, we will see in
Theorem~\ref{thm:unitarytrick} that even in this case it is possible
to choose the coefficients $\alpha_g$ such that the associated
group-circulant is unitary.
\smallskip

The notion of circulant matrices is based on ordinary representations
of a given finite group $G$. It is possible to generalize the concepts
presented in this section to projective representations. In doing so,
a greater flexibility in forming linear combinations can be
achieved. This will be studied in detail in Section
\ref{projCirculants}. \medskip

%%%%%%%%%%%%%%%%%%%%%%%%%%%%%%%%%%%%%%%%%%%%%%%%%%%%%%%%%%%%
%
% Section: Case-Operators
%
%%%%%%%%%%%%%%%%%%%%%%%%%%%%%%%%%%%%%%%%%%%%%%%%%%%%%%%%%%%%

\goodbreak
\section{Case-Operators}\label{group-indexed}%
Let $G$ be a finite group, and denote by $D\colon G\rightarrow
{\mathcal{U}}(2^m)$ an ordinary matrix representation of $G$ acting
by unitary matrices on a system of $m$ quantum bits. Our goal is to derive 
an efficient implementation of a block
diagonal matrix $D^\oplus$ containing the representing matrices $D(g)$,
$$ D^\oplus = \diag(D(g)\colon g\in G).$$ This will be an essential
step in creating a linear combination of these matrices. We need an
efficient implementation of this block diagonal matrix, and a suitable
encoding of the group elements will allow us to find such an
implementation.

For simplicity, we assume that $G$ is a $2$-group, that is, $|G|=2^n$
for some integer $n\ge 1$, but the ideas easily generalize to
arbitrary solvable groups. It is possible to find a
composition series $E = G_0 \lhd G_1 \lhd \ldots \lhd G_n = G$ of the
group~$G$ such that the quotient group $G_{i+1}/G_i$ contains exactly two elements, $G_{i+1}/G_i\cong \Z_2$,
see~\cite{Huppert:79}. A {\em transversal}\/ of $G$ is a sequence of
elements $T=(t_1, \ldots, t_n)$ such that $t_i \in G_i$, and the quotient group $G_i/G_{i-1}$ is generated by the image of the element $t_i\in G_i$, 
\[ \langle \overline{t_i} \rangle = G_i/G_{i-1}\quad
\mbox{for}\quad i=1,\ldots,n.
\]
The essence of this somewhat technical construction is that we obtain a unique presentation of each element $g\in G$ in the form 
\begin{equation}\label{represent}
 g = t_1^{a_1} \cdot \ldots \cdot t_n^{a_n} \quad 
\mbox{with}\quad a_i \in \{0,1\}.
\end{equation}
This allows to ``address'' each group element by a binary string of
$n$ bits. Abusing notation, we identify the element $g$ with its
exponent vector $(a_1,\dots, a_n)$, and we write $D(a_1,\dots, a_n)$ to denote the matrix 
$$ D(a_1,\dots, a_n) = D(t_1^{a_1}\cdots t_n^{a_n}).$$
Let $D^\oplus_T \in {\cal U}(2^{n+m})$ denote 
the block diagonal matrix
\[ D^\oplus_T = \left(
\begin{array}{cccc}
{D{(0,\dots,0)}} & & & \\
& {D{(0,\dots,0,1)}} & & \\
& & \ddots & \\
& & & {D{(1, \dots, 1)}}
\end{array}
\right).
\]
This block diagonal matrix contains the representing matrix of each
group element $g=t^{a_1}_1\cdots t_n^{a_n}$.  We need only an
implementation of the matrices $D(t_1), \dots, D(t_n)$, because the
representing matrices satisfy the relation $D(t_i)D(t_j)=D(t_it_j)$.
We will conditionally apply these matrices on the system of $m$ quntum
bits. We have $n$ control bits, one for each matrix $D(t_i)$. 

We need a lemma which allows us to give an estimate of the complexity
of our implementation.

\begin{lemma}
Let $U$ be an elementary gate, i.e., an element of ${\cal G}$. 
Then the conditional gate $\Lambda_1(U)$ can
be implemented using at most $14$ elementary gates. 
\end{lemma}
\smallskip
{\bf Proof.} If $U$ is a single-qubit gate, then $\Lambda_1(U)$ can be
implemented with at most six elementary gates~\cite{BBC+:95}. If $U$ is a
controlled-not gate, then $\Lambda_1(U)$ is a Toffoli gate, which can
be implemented with 14 elementary gates~\cite{DiVincenzo:98}.~\qed
\smallskip

We now state the main theorem of this section, which gives an upper
bound on the complexity of the case operator $U^\oplus_T$:

\begin{theorem}\label{blockdecomp}
Let $G$ be a finite group of order $2^n$ with 
a unitary matrix representation $D\colon G\rightarrow {\mathcal{U}}(2^m)$. 
Let $T=(t_1, \ldots, t_n)$ denote a transversal of $G$.
If $c_T =
\max_{t \in T} \kappa(U_t)$ is the maximum number of operations
necessary to realize one of the matrices $D(t)$, $t\in T$,  
then the block diagonal matrix  $D_T^\oplus$ can be realized with at most $\kappa(D^\oplus_T) \leq
14\, n\, c_T$ elementary operations. 
\end{theorem}

% Zerlegte Twiddles
% -----------------
\begin{figure}[htb]
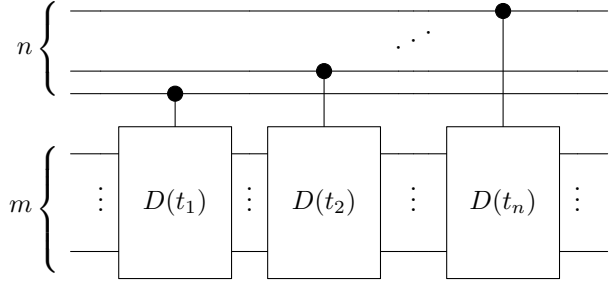

\begin{center}
\begin{emp}(50,50)
  setunit(2mm);
  qubits(5);
  QCycoord[1] := QCycoord[1]+0.5cm;
  QCycoord[2] := QCycoord[2]+0.5cm;
  QCycoord[3] := QCycoord[3];

  label.lft(btex $n$ etex, (QCxcoord,QCycoord[3]+3mm));
  label.lft(btex $m$ etex, (QCxcoord,QCycoord[0]+0.6cm));
  QCxcoord := QCxcoord+4mm;

  label.lft(btex $\left\{\rule{0mm}{0.7cm}\right.$ etex,(QCxcoord, QCycoord[2]+0.6cm));
label.lft(btex $\left\{\rule{0mm}{1cm}\right.$ etex,(QCxcoord, QCycoord[0]+0.6cm));

  wires(4mm);
  label(btex $\vdots$ etex, (QCxcoord, QCycoord[0]+0.8cm));
  circuit(1.5cm)(icnd 2, gpos 0,1, btex $D({t_1})$ etex);
  label(btex $\vdots$ etex, (QCxcoord, QCycoord[0]+0.8cm));
  circuit(1.5cm)(icnd 3, gpos 0,1, btex $D({t_2})$ etex);
  label(btex $\cdot$ etex, (QCxcoord, QCycoord[3]+3mm));
  label(btex $\cdot$ etex, (QCxcoord+2mm, QCycoord[3]+4mm));
  label(btex $\cdot$ etex, (QCxcoord+4mm, QCycoord[3]+5mm));
  wires(2mm);
  label(btex $\vdots$ etex, (QCxcoord, QCycoord[0]+0.8cm));
  wires(2mm);
  circuit(1.5cm)(icnd 4, gpos 0,1, btex $D({t_n})$ etex);
  label(btex $\vdots$ etex, (QCxcoord, QCycoord[0]+0.8cm));
  wires(4mm);

\end{emp}
\end{center}
\caption{\label{twiddledecomp}Quantum circuit for a
transversal case operator}
\end{figure}

\noindent
{\bf Proof.} We observe that due to binary expansion of the exponent
vectors the operation $U^\oplus_T$ can be implemented as in
Figure~\ref{twiddledecomp}. The statement concerning the number of
gates of this factorization follows immediately from the previous
lemma.~\qed
\smallskip

A familiar example is given by the additive cyclic group
$\Z/2^n\Z$. Assume that this group is represented by $D(g)=U^g$, where
$g\in \Z/2^n\Z$ and $U$ is some unitary matrix satisfying
$U^{2^n}=\onemat$. A composition series is given by the subgroups
$G_i= 2^{n-i}\Z/2^n\Z$.  A transversal for the group $\Z/2^n\Z$ is, for
instance, given by the group elements $t_i=2^{n-i}$, that is, $T=(2^{n-1},
2^{n-2},\dots, 2,1)$.  The implementation described in the previous
theorem realizes the powers $U^{2^i}$. An arbitrary power $U^g$ is
realized by setting the $n$ control bits according to the binary
expansion $ g= \sum g_i 2^{n-i}$, with $g_i\in \{0,1\}$.

%%%%%%%%%%%%%%%%%%%%%%%%%%%%%%%%%%%%%%%%%%%%%%%%%%%%%%%%%%%%
%
% Section: The New Design Principle
%
%%%%%%%%%%%%%%%%%%%%%%%%%%%%%%%%%%%%%%%%%%%%%%%%%%%%%%%%%%%%

\goodbreak
\section{The Design Principle}\label{design}
Suppose that we want to realize a unitary matrix $A\in
{\mathcal{U}}(2^m)$ by a quantum circuit. We assume that some unitary
matrices $D(g)\in {\mathcal{U}}(2^m)$ with efficient quantum
circuits are known to us. Familiar examples are discrete Fourier
transforms, permutation matrices, and so on.  Suppose that some of
the matrices $D(g)$ generate a finite dimensional group algebra
containing the matrix $A$, then, simply put, a quantum circuit can be
found for $A$. The following theorem describes how this can be
accomplished.  To ease the presentation, we do not state the theorem
in its most general form. The more technical generalizations will be
discussed in the subsequent sections.
\smallskip

\begin{theorem}\label{completeCircuit}
Let $G$ be a finite group of order $2^n$, and denote by $T=(t_1,
\ldots, t_n)$ a transversal of $G$, that is, each element $g\in G$ can
be uniquely represented in the form $g = t_1^{a_1} \cdot \ldots \cdot
t_n^{a_n}$, where $a_i \in \{0,1\}$.  Let $D: G \rightarrow {\cal
U}(2^m)$ be a unitary representation of $G$ such that the images
$\{D(g) : g \in G\}$ form a set of linearly independent unitary
operations. Suppose that $A\in {\cal U}(2^m)$ is
a linear combination of the representing matrices $D(g)$, 
\[ A = \sum_{g \in G} \alpha_g D(g),\]
with coefficients $\alpha_g \in \C$. If\/ $C_A =
\zirc_G(\ket{\alpha})$ denotes the associated group-circulant,
with elements ordered according to the choice of the transversal $T$, 
then the matrix $A$ is realized by the circuit given in Figure~\ref{generic}.
\end{theorem}
% Generischer Schaltkreis
% -----------------------
\begin{figure}[htb]
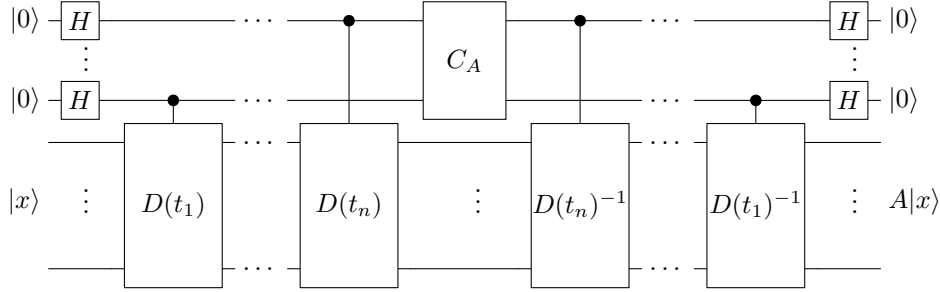

\begin{center}
\begin{emp}(50,50)
  setunit(1.4mm);
  qubits(6);
  dropwire(1,2);
  QCycoord[3] := QCycoord[3] + 5mm;  
  label.lft(btex $|0\rangle$ etex, (QCxcoord, QCycoord[3]));  
  label.lft(btex $|0\rangle$ etex, (QCxcoord, QCycoord[2]));
  label.lft(btex $|x\rangle$ etex, (QCxcoord, QCycoord[0]+9mm));

  label(btex $\vdots$ etex, (QCxcoord+5mm, QCycoord[2]+6.5mm));
  label(btex $\vdots$ etex, (QCxcoord+5mm, QCycoord[0]+10mm));
  gate(gpos 2, btex $H$ etex, 3, btex $H$ etex);
  circuit(1.3cm)(icnd 2, gpos 0,1, btex  $D(t_1)$ etex);
  label.rt(btex $\dots$ etex, (QCxcoord, QCycoord[0]));
  label.rt(btex $\dots$ etex, (QCxcoord, QCycoord[1]));
  label.rt(btex $\dots$ etex, (QCxcoord, QCycoord[2]));
  label.rt(btex $\dots$ etex, (QCxcoord, QCycoord[3]));
  QCxcoord := QCxcoord + 7mm;
  circuit(1.3cm)(icnd 3, gpos 0,1, btex  $D(t_n)$ etex);
  label(btex $\vdots$ etex, (QCxcoord+9mm, QCycoord[0]+10mm));
  circuit(1.1cm)(gpos 2,3, btex  $C_A$ etex);
  
  circuit(1.3cm)(icnd 3, gpos 0,1, btex  $D(t_n)^{-1}$ etex);
 label.rt(btex $\dots$ etex, (QCxcoord, QCycoord[0]));
  label.rt(btex $\dots$ etex, (QCxcoord, QCycoord[1]));
  label.rt(btex $\dots$ etex, (QCxcoord, QCycoord[2]));
  label.rt(btex $\dots$ etex, (QCxcoord, QCycoord[3]));
  QCxcoord := QCxcoord + 7mm;

  circuit(1.3cm)(icnd 2, gpos 0,1, btex  $D(t_1)^{-1}$ etex);
  label(btex $\vdots$ etex, (QCxcoord+5mm, QCycoord[2]+6.5mm));
  label(btex $\vdots$ etex, (QCxcoord+5mm, QCycoord[0]+10mm));
  gate(gpos 2, btex $H$ etex, 3, btex $H$ etex);
  label.rt(btex $|0\rangle$ etex, (QCxcoord, QCycoord[3]));  
  label.rt(btex $|0\rangle$ etex, (QCxcoord, QCycoord[2]));
  label.rt(btex $A|x\rangle$ etex, (QCxcoord, QCycoord[0]+9mm));

\end{emp}
\end{center}
\caption{\label{generic} 
Quantum circuit implementing a linear combination}
\end{figure}

\noindent
{\bf Proof.}  Note that by the choice of the transversal $T$ the
ordering of the elements of $G$ is fixed. We define $H$ to be the
transformation $H := H_2 \otimes \ldots \otimes H_2 \otimes
\onemat_{2^n}$, where the first $n$ factors are equal to the Hadamard
transform $H_2$.  The transformation $H$ corresponds to the leftmost
and to the rightmost transformation in
Figure~\ref{generic}. Furthermore, we define $C_A :=
\zirc(\ket{\alpha}) \otimes \onemat_{2^n}$ and ${\cal D} :=
\diag(D(g_1), \ldots, D(g_{2^n}))$.  Observe that ${\cal D}$, $C_A$,
and ${\cal D}^{-1}$ are the remaining transformations in Figure
\ref{generic}.  The circuits for ${\cal D}$ and ${\cal D}^{-1}$ are
shown in factorized form, hereby exploiting the group-structure of the
case-operator. We obtain the factorization of ${\cal D}$ as in Figure
\ref{twiddledecomp}.  To verify that the circuit indeed computes
$\ket{0} \ket{x} \mapsto \ket{0}A\ket{x}$, we first consider the
matrix identity
\begin{equation}\label{blockMat}
{\cal D}^{-1} \cdot C_A \cdot {\cal D} = \left(\begin{array}{cccc}
    \alpha_{g_1} \onemat_{2^n} & \alpha_{g_2} D(g_2) & \cdots & \alpha_{g_{2^n}} D(g_{2^n})\\
    \alpha_{g_2^{-1}} D(g_2^{-1}) & \alpha_{g_1} \onemat_{2^n} & \cdots & \vdots \\
    \vdots  &\ddots&\ddots& \vdots\\
    \alpha_{g_{2^n}^{-1}} D(g_{2^n}^{-1}) & \cdots & \cdots & \alpha_{g_1}
    \onemat_{2^n}
\end{array}\right).
\end{equation}
To be more precise, the entry at position $(i,j)$ of this
block-structured matrix is equal to $\alpha_{g_i^{-1} g_j} D(g_i^{-1})
D(g_j) = \alpha_{g_i^{-1} g_j} D(g_i^{-1} g_j)$. This means that each
row of blocks of (\ref{blockMat}) contains the set of matrices $\{
\alpha_g D(g)\,\colon\, g \in G \}$ in some permuted order.  The same
holds for the columns of this matrix.  Hence we can conclude that the
first row of the matrix $H^{-1} {\cal D}^{-1} \cdot C_A \cdot {\cal
D}$ is given by $\frac{1}{\sqrt{2^n}} (\sum_g \alpha_g D(g), \ldots,
\sum_g \alpha_g D(g)) = \frac{1}{\sqrt{2^n}} (A, \ldots, A)$. Hence
applying $H$ to the columns of this matrix will produce
\begin{equation}\label{blockfact}
H^{-1} \cdot {\cal D}^{-1} \cdot C_A \cdot {\cal D} \cdot H = 
\left(\begin{array}{cc}
A & \zeromat \\
\zeromat & U
\end{array}\right),
\end{equation}
with some unitary matrix $U \in {\cal U}(2^{n+m}{-}2^m)$ and
zero-matrices of the appropriate sizes. Note that the entries in the same
rows resp. columns as $A$ must vanish, since $A$ as well as the other
operations used in (\ref{blockfact}) are unitary.  \qed
\smallskip

\textit{Remark.} Note that the assumptions in the theorem can be
considerably relaxed. The restriction to 2-groups is not necessary. In
fact, the implementation of case operators can be extended to
arbitrary solvable groups. Moreover, we will show in the next section
that the representing matrices $D(g)$ do not need to be linearly
independent; one can always find suitable unitary group circulants
$C_A$. Finally, it is not necessary that the representing matrices
form an ordinary representation; the extension to projective
representations is discussed in Section~\ref{projCirculants}.
\medskip

Note that the cost of implementing a circuit for $A$ is determined by
the cost of the transversal elements $\kappa(D_{t_i})$, $1\le i\le n$,
and by the cost of the group circulant $\kappa(C_A)$.  If the group
$G$ is of small order, say the order is bounded by $c$, then the
efficiency of the implementation of a transformation which has been
decomposed according to Theorem \ref{completeCircuit} depends only on
the complexity of the transformations $\kappa(D_{t_i})$. A family of
representations with this property will be studied in
Section~\ref{fractional}.

%%%%%%%%%%%%%%%%%%%%%%%%%%%%%%%%%%%%%%%%%%%%%%%%%%%%%%%%%%%%
%
% Section: Generalization
%
%%%%%%%%%%%%%%%%%%%%%%%%%%%%%%%%%%%%%%%%%%%%%%%%%%%%%%%%%%%%

\section{Generalization}\label{sec:generalization} 
The theorem in the previous section assumed that the representing
matrices $D(g)$ of the group $G$ are linearly independent.
The next theorem shows that one can drop this assumption entirely. 

\begin{theorem}\label{thm:unitarytrick}  
Let $D\colon G\rightarrow {\mathcal{U}}(2^n)$ be an ordinary 
representation of a finite group $G$. If a unitary matrix
$A$ can be expressed as a linear combination 
$$ A = \sum_{g\in G} \alpha_g D(g),\qquad \alpha_g\in \C, $$
then the coefficients $\alpha_g$ can be chosen such that the 
associated group circulant is unitary. 
\end{theorem}
\smallskip
\noindent{\bf Proof.}  A finite group has a finite number of
non-equivalent irreducible representations.  Let $D^{(k)}\colon
G\rightarrow {\mathcal{U}}(n_k)$, $k\in I$, be a representative set of
the non-equivalent irreducible unitary representations of the finite
group~$G$.

Let $D_{ij}^{k}\colon G\rightarrow \C$ be the complex-valued function
on $G$, which is determined by the value of the $(i,j)$-coefficient of
the representing matrix $D^{(k)}$. We obtain $\sum_{k\in I} n_i^2$
functions in this way.  It was shown by Schur that these functions are
orthogonal, 
$$ \langle D_{ij}^k | D_{i'j'}^{k'}\rangle = 0, $$ 
unless $i=i'$, $j=j'$, and $k=k'$; see~\cite[Section~2.2]{serre96}.

Denote by
$$J = \{ k\in I\colon D^{(k)} \mbox{ is a constituent of the
representation } D\}.$$ 
Let $E$ denote the direct sum of the irreducible representations, 
which are not contained in~$D$, that is,  
$$ E(g) = \bigoplus_{k \in I- J} D^{(k)}(g).$$ 
We have 
$$ A\oplus I = \sum_{g\in G} \alpha_g (D(g)\oplus E(g))$$ for some
$\alpha_g\in \C$. Indeed, comparing coefficients yields a system of
$|G|$ linear equations. This system of equations can be solved, 
since the coefficient functions are orthogonal. The circulant corresponding to
the coefficients $\alpha_g$ is unitary by Theorem~\ref{procUnit}.
Ignoring the representing matrices $E(g)$, we obtain $A$ as a linear
combination of the representing matrices $D(g)$, as claimed.~\qed

%%%%%%%%%%%%%%%%%%%%%%%%%%%%%%%%%%%%%%%%%%%%%%%%%%%%%%%%%%%%
%
% Section: Extension to Projective Circulants
%
%%%%%%%%%%%%%%%%%%%%%%%%%%%%%%%%%%%%%%%%%%%%%%%%%%%%%%%%%%%%

\section{Extension to Projective Circulants}
\label{projCirculants}

We have assumed in Theorem \ref{completeCircuit} that $A$ is obtained
as a linear combination of matrices $D(g)$, which form an ordinary
representation of a finite group $G$. It turns out that the quantum
circuit used for the implementation of $A$ can also be used, with a
minor modification, for projective representations.  We recall a few
basic facts about projective representations and then give the
appropriate generalization of the circulant matrices introduced in
Section \ref{circulant}.

Note that projective representations have been used in quantum
information theory. They for instance turn out to be the adequate
formalism to describe a class of unitary error bases~\cite{KR:2002a}.

Let $D\colon G\rightarrow U(n)$ be a projective unitary representation
of a finite group $G$ with factor set $\omega$. In other words,
$\omega$ is a function from $G\times G$ to the nonzero complex numbers
$\C^\times$ such that
\begin{equation}\label{mult} 
D(g)D(h) = \omega(g,h)D(gh)
\end{equation}
holds for all $g,h\in G$. The associativity of 
matrix multiplication implies the relations
\begin{equation}\label{cocycle} 
\omega(x,y)\omega(xy,z) = \omega(x,yz)\omega(y,z)
\end{equation}
for all $x,y,z\in G$. This shows that $\omega$ is a $2$-cocycle
of the group $G$ with trivial action on $\C^\times$. 

We assume that the neutral element $1$ of the group $G$ is
represented by the identity matrix $D(1)=1_n$, which implies 
\begin{equation}\label{normalization} 
\omega(g,1)=\omega(1,g)=1\qquad \mbox{for all}\quad g\in G.
\end{equation}
The values of the factor system $\omega$ are of modulus 1, 
since the representation matrices $D(g)$ are unitary. This shows, in
particular, the relations
\begin{equation}\label{unit}
\overline{\omega(g,h)}=\omega(g,h)^{-1}\qquad \mbox{for}\quad g,h\in G.
\end{equation}

Let $\alpha \in \C^{|G|}$ be a vector which is labeled by the elements
of $G$. We define a {\em projective group circulant} for $(G,\omega)$
with respect to $\alpha$ to be the matrix
\[ \zirc_{G,\omega}(\alpha) =\left(\,\omega(g^{-1},h)^{-1}\,\alpha_{g^{-1}h}\,\right)_{g,h\in G}.
\]
Projective circulants have been introduced by I.~Schur
\cite{Schur:09}. We show that in the analog situation to Theorem
\ref{completeCircuit} the associated projective circulants are
unitary.

\begin{theorem}\label{procUnit}\
Let $D\colon\, G\rightarrow {\cal U}(n)$ be an $n$-dimensional unitary
projective representation of a finite group $G$.
Suppose that the operators $\{ D(g)\colon g\in G\}$ 
are 
linearly independent. If a unitary matrix $A\in {\cal U}(n)$ 
can be expressed as a linear combination 
\[ A = \sum_{g \in G} \alpha_g D(g),\qquad \mbox{with} 
\quad\alpha_g\in \C,\]
then the projective group circulant 
$\zirc_{G,\omega}(\alpha)$ of the 
coefficients $\alpha_{g}$ is unitary.
\end{theorem}

\noindent
{\bf Proof.}$\;$ 
It suffices to show that the rows of the projective group circulant
$\zirc_{G,\omega}(\alpha)$ are pairwise 
orthogonal and of unit length, or more explicitly that 
\begin{equation}\label{ortho} 
\sum_{h\in G} (\omega(g,h)\overline{\omega(k,h)})^{-1} \alpha_{gh}
\overline{\alpha_{kh}}=\delta_{g,k}
\end{equation}
holds for all $g,k\in G$. We will show that these orthogonality
relations can be derived 
from the matrix identity
$AA^\dagger D(k)=D(k)$.

Multiplying $A, A^\dagger$ and $D(k)$ yields
\begin{eqnarray*}
A A^\dagger D(k) & = & \Big(\sum_{g \in G} \alpha_g D_g \Big)
\Big(\sum_{h \in G} \overline{\alpha_h} D(h)^\dagger \Big) D(k) \\
& = & \sum_{g \in G} \sum_{h \in G} \alpha_{g} 
\overline{\alpha_h} D(g) D(h)^\dagger D(k).
\end{eqnarray*}
Notice that the multiplication rules (\ref{mult}) imply
$$ D(g) D(h)^\dagger D(k) = 
\frac{\omega(g,h^{-1})\omega(gh^{-1},k)}{\omega(h,h^{-1})} D(gh^{-1}k).
$$
Therefore, $AA^\dagger D(k)=D(k)$ can be expressed as
\begin{eqnarray*}
D(k)
&= & \sum_{g \in G} \sum_{h \in G} \alpha_{g} \overline{\alpha_h}
\frac{\omega(g,h^{-1})\omega(gh^{-1},k)}{\omega(h,h^{-1})} D(gh^{-1}k)\\
&\stackrel{\ell=gh^{-1}}{=}& \sum_{\ell \in G} \Big(\sum_{h \in G}
\alpha_{\ell h} \overline{\alpha_h}
\frac{\omega(\ell h,h^{-1})\omega(\ell,k)}{\omega(h,h^{-1})}
\Big) D(\ell k).
\end{eqnarray*}
Setting $x:=\ell $, $y:=h $ and $z:=h^{-1}$ in (\ref{cocycle}) shows 
the identity
$$ \omega(\ell h,h^{-1})\omega(h,h^{-1})^{-1} = 
\omega(\ell,h)^{-1},
$$ 
which allows to simplify the previous expression for $D(k)$ to 
\begin{eqnarray*} 
D(k) &=& \sum_{\ell \in G} \Big(\sum_{h \in G}
\alpha_{\ell h} \overline{\alpha_h}
\frac{\omega(\ell, k)}{\omega(\ell,h)}
\Big) D(\ell k) \\
 &=& \sum_{\ell \in G} \Big(\sum_{h \in G}
\alpha_{\ell kh} \overline{\alpha_{kh}}
\frac{\omega(\ell, k)}{\omega(\ell,kh)}
\Big) D(\ell k).
\end{eqnarray*}
The substitution $g=\ell k$ yields
$$ D(k) = 
\sum_{g \in G} \Big(\sum_{h \in G}
\alpha_{g h} \overline{\alpha_{kh}}
\frac{\omega(gk^{-1},k)}{\omega(gk^{-1},kh)}
\Big) D(g).
$$
Setting 
$x:= gk^{-1}$, $y:=k$, and $z:=h$ in the cocycle
relation (\ref{cocycle}) shows the identity
$$
\frac{\omega(gk^{-1},k)}{ \omega(gk^{-1},kh)} = \frac{\omega(k,h)}{\omega(g,h)}.
$$
This allows to write $D(k)$ in the form 
$$D(k) =  
\sum_{g \in G} \Big(
\sum_{h\in G} \frac{\omega(k,h)}{\omega(g,h)}
\alpha_{gh}
\overline{\alpha_{kh}}\Big) D({g})$$
Comparing coefficients on both sides yields
\nix{$$
\sum_{h\in G} \frac{\omega(k,k^{-1}h)}{\omega(g,k^{-1}h)}
\alpha_{gk^{-1}h}
\overline{\alpha_{h}}=\delta_{g,k}
\quad\mbox{for all}\quad g,k\in G.
$$
If we substitute $hk$ for $k$, then we obtain
}
$$ \sum_{h\in G} \frac{\omega(k,h)}{\omega(g,h)}
\alpha_{gh}
\overline{\alpha_{kh}}=\delta_{g,k}\quad\mbox{for all}\quad g,k\in
G.$$
Using (\ref{unit}), this shows that the 
orthogonality relations (\ref{ortho}) hold. Thus, the projective group
circulant $\zirc_{G,\omega}(\alpha)$ 
is indeed unitary, as claimed.~\qed

We remark that a theorem analogous to Theorem \ref{completeCircuit}
holds in the situation where $D$ is a projective representation of a
finite group $G$. In this case the matrices $D_{t_i}$ have to be
replaced by a suitably rescaled transversal and the circulant matrix
$C_A$ has to be replaced by the corresponding projective
circulant. Theorem \ref{procUnit} guarantees that the latter matrix is
unitary.

Using projective representations a greater flexibility can be
achieved. In Section \ref{teleport} we give an example for a
projective representation of the group $\Z_2 \times \Z_2$ which is
given by the Pauli matrices.

%%%%%%%%%%%%%%%%%%%%%%%%%%%%%%%%%%%%%%%%%%%%%%%%%%%%%%%%%%%%
%
% Section: Connection with Kitaevs algorithm
%
%%%%%%%%%%%%%%%%%%%%%%%%%%%%%%%%%%%%%%%%%%%%%%%%%%%%%%%%%%%%

\section{Relation to Kitaev's algorithm}
\label{kitaevRelation}

Kitaev presented in \cite{Kitaev:95} and \cite{Kitaev:97}
a quantum circuit that allows to estimate an eigenvalue of
a unitary matrix provided that the corresponding eigenstate is given; see  
also \cite{CEMM:98} and \cite{Jozsa:98} for further descriptions
of this scenario. The phase estimation provides a unified framework
for Shor's algorithm \cite{Shor:94} and the algorithms for abelian
stabilizers~\cite{Kitaev:95,Grigoriev:96}.
In the following we give a brief account of this method. Starting from
a unitary transformation $U$ on $n$ qubits and an eigenvector
$\ket{\psi_\lambda}$, we want to generate an estimate of
$\lambda$. The precision of this approximation is controlled by the
number $k$ of digits we want to compute in the binary expansion
$\lambda = 0.\,b_1\,b_2\,b_3\, \ldots$ of $\lambda$, i.\,e., $b_i \in \F_2$. 

If we assume that, in addition to $U$ and $\ket{\psi_\lambda}$, we are
given efficient quantum circuits implementing $\Lambda_1(U^{2^j})$ for
$j=0, \ldots, k-1$, then it is possible to accomplish the task of
approximating $\lambda$ by means of an efficient quantum circuit. This
circuit, which is given in Figure \ref{kitaevCircuit}, consists of
three parts.

Reading from left to right, we have a quantum
circuit acting on two registers: the first holds the eigenstate
$\ket{\psi_\lambda}$ and the second, which ultimately will contain the
approximation $|\tilde{\lambda}\rangle$, is initialized with the
$\ket{0}$ state. In a first step an equal-weighted superposition of
all binary strings on the second register is generated by application
of a Hadamard transform to each wire. Then the transformation
$U^\oplus := \ket{i}\ket{x} \mapsto \ket{i} U^i \ket{x}$ is performed for $i=0,
\ldots, 2^k - 1$. Note that if $U$ has finite order $2^k$,
then $U^\oplus$ is a case-operator (in the sense of
Section \ref{group-indexed}) for the cyclic group $Z_{2^k}$.
In a third step an inverse Fourier transform $\DFT_{2^k}^{-1}$ is
computed giving the best $k$-bit approximation of $\lambda$ which is
stored in the second register~\cite{CEMM:98}. 

% Quantum circuit for eigenvalue estimation
% -----------------------------------------
\nix{\begin{figure}
\input{pic/kitaev.pic}
\caption[]{Quantum circuit for eigenvalue
estimation}
\end{figure}
}
\begin{figure}
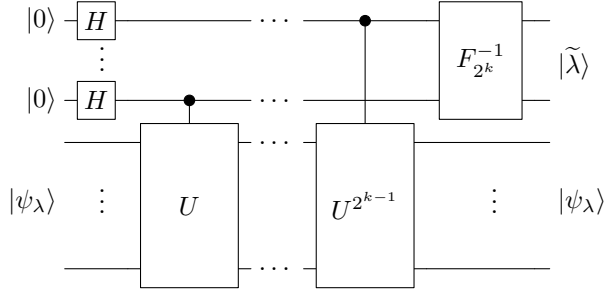

\begin{center}
\begin{emp}(50,50)
  setunit(1.4mm);
  qubits(6);
  dropwire(1,2);
  QCycoord[3] := QCycoord[3] + 5mm;  
  label.lft(btex $|0\rangle$ etex, (QCxcoord, QCycoord[3]));  
  label.lft(btex $|0\rangle$ etex, (QCxcoord, QCycoord[2]));
  label.lft(btex $|\psi_\lambda\rangle$ etex, (QCxcoord, QCycoord[0]+9mm));

  label(btex $\vdots$ etex, (QCxcoord+5mm, QCycoord[2]+6.5mm));
  label(btex $\vdots$ etex, (QCxcoord+5mm, QCycoord[0]+10mm));
  gate(gpos 2, btex $H$ etex, 3, btex $H$ etex);
  circuit(1.3cm)(icnd 2, gpos 0,1, btex  $U$ etex);
  label.rt(btex $\dots$ etex, (QCxcoord, QCycoord[0]));
  label.rt(btex $\dots$ etex, (QCxcoord, QCycoord[1]));
  label.rt(btex $\dots$ etex, (QCxcoord, QCycoord[2]));
  label.rt(btex $\dots$ etex, (QCxcoord, QCycoord[3]));
  QCxcoord := QCxcoord + 7mm;
  circuit(1.3cm)(icnd 3, gpos 0,1, btex  $U^{2^{k-1}}$ etex);
  label(btex $\vdots$ etex, (QCxcoord+9mm, QCycoord[0]+10mm));
  circuit(1.1cm)(gpos 2,3, btex  $F_{2^{k}}^{-1}$ etex);
  wires(2mm);

  label.rt(btex $|\widetilde{\lambda}\rangle$ etex, (QCxcoord, QCycoord[2]+5mm));
  label.rt(btex $|\psi_\lambda\rangle$ etex, (QCxcoord, QCycoord[0]+9mm));

\end{emp}
\end{center}
\caption{\label{kitaevCircuit} Quantum circuit for eigenvalue
estimation}
\end{figure}

The connection between the algorithm for phase estimation and the
circuit given in Figure \ref{generic} is established as follows. For
the special case of $G=\Z/{2^k}\Z$ generated by a unitary transformation
$U\in {\cal U}(2^n)$ of order $2^k$ we first apply the circuit given
in Figure \ref{kitaevCircuit} to each element of a basis $\{
\ket{\psi_i} : i = 1,\ldots,2^k \}$ of eigenvectors of $U$ in order to
obtain the (exact) eigenvalues $\lambda_i$ in the second
register. Note that there are at most $2^k$ different eigenvalues of
$U$. We then perform the scalar multiplication $\lambda_i \mapsto
\alpha_i \lambda_i$ for certain $\alpha_i \in {\cal U}(1)$ and
all $i=1,\ldots,2^k$. Finally we run the circuit given in Figure
\ref{kitaevCircuit} backwards and observe that
\[ 
\DFT_{2^k}^{-1} \cdot {\rm diag}(\alpha_1, \ldots, \alpha_{2^k})
\cdot \DFT_{2^k} = \zirc_{\Z/{2^k}\Z}(\beta_1, \ldots, \beta_{2^k}),
\]
where the vector $\ket{\beta} = (\beta_1, \ldots, \beta_{2^k}) \in
\C^{2^k}$ is given by $\ket{\beta} = F_{2^k} \ket{\alpha}$.
Here $F_N = [e^{2 \pi i k l/N}]_{k,l=0,\ldots,N-1}$ denotes the
(unnormalized) discrete Fourier transform.

Note, that the method presented in Section \ref{design} is more
general than this twofold application of the circuit for eigenvalue
estimation as it allows to work with representations of arbitrary
finite groups.

%%%%%%%%%%%%%%%%%%%%%%%%%%%%%%%%%%%%%%%%%%%%%%%%%%%%%%%%%%%%
%
% Section: Examples 
%
%%%%%%%%%%%%%%%%%%%%%%%%%%%%%%%%%%%%%%%%%%%%%%%%%%%%%%%%%%%%

\section{Examples}
\label{examples}

The decomposition method introduced in the previous sections is
demonstrated by means of the {\em Hartley transforms} which have been
introduced in Section \ref{motivation} and {\em fractional Fourier
transforms} (cf. \cite{BM:96,CKO:2000}) which are a class of unitary
transformations used in classical signal processing. Using the method
of linear combinations of unitary operations we show how to compute
them efficiently on a quantum computer. Finally, we show that the
quantum circuit for teleportation of a qubit can be interpreted with the
help of this method.

%%%%%%%%%%%%%%%%%%%%%%%%%%%%%%%%%%%%%%%%%%%%%%%%%%%%%%%%%%%%
%
% Subsection: Hartley Transforms Revisited
%
%%%%%%%%%%%%%%%%%%%%%%%%%%%%%%%%%%%%%%%%%%%%%%%%%%%%%%%%%%%%

\subsection{Hartley Transforms Revisited}
\label{hartley}

The efficient quantum circuit shown in Figure \ref{hartleyCirc} of
Section \ref{motivation} can be recast in terms of Theorem
\ref{completeCircuit}. First recall that the identity
\[ A_N = \alpha \; F_N 
    +  \beta \; F_N^3
\] 
with $\alpha := \left(\frac{1-i}{2}\right)$ and $\beta :=
\left(\frac{1+i}{2}\right)$ shows that the Hartley transform $A_N$ is
a linear combination of the powers $\onemat_N, F_N, F_N^2, F_N^3$ of
the discrete Fourier transform $F_N$. We can simplify this to obtain
$A_N = F_N \tilde{A}_N$, where $\tilde{A}_N$ is defined as
$\tilde{A}_N := \alpha \onemat_N + \beta F_N^2$. Since $F_N^2$ is an
involution, we can apply apply Theorem \ref{completeCircuit} in the
special situation where $|G|=2$. Hence the circulant $C_{\tilde{A}_N}$
is in this case the $\Z/2\Z$ circulant matrix 
\[ 
R := \frac{1}{2}
\left( \begin{array}{rr} 1-i & 1+i \\ 1+i & 1-i \end{array} \right).
\]
Since for $N\geq 5$ the matrices $\onemat_N, F_N, F_N^2, F_N^3$ are
linearly independent, we can use Theorem \ref{completeCircuit} to
conclude that $R$ has to be unitary. Hence we can implement
$\tilde{A}_N$ using one auxiliary qubit with a quantum circuit as in
Figure \ref{generic}. Combining the circuits for $F_N$ and for
$\tilde{A}_N$ we finally obtain the factorization of $A_N$ shown in
Figure \ref{hartleyCirc} of Section \ref{motivation}.

%%%%%%%%%%%%%%%%%%%%%%%%%%%%%%%%%%%%%%%%%%%%%%%%%%%%%%%%%%%%
%
% Subsection: Fractional Fourier Transforms
%
%%%%%%%%%%%%%%%%%%%%%%%%%%%%%%%%%%%%%%%%%%%%%%%%%%%%%%%%%%%%

\subsection{Fractional Fourier Transforms}
\label{fractional}

\noindent
A matrix $A$ having the property $A^\alpha = B$ with $\alpha \in \R$
is called an $\alpha$-th root of $B$ (where in general this root is
not uniquely determined). In case of the discrete Fourier transform
$F_N$ we can use the property that $F_N^4 = \onemat_N$ to define an
$\alpha$-th root of $F_N$ via
\begin{equation}\label{fractFact}
 F_{N, \alpha} := a_0(\alpha) \cdot \onemat_N + \ldots + 
a_3(\alpha) \cdot F_N^3,
\end{equation}
where the coefficients $a_i(\alpha)$ for $i=0,\ldots,3$ are defined by
\[
\renewcommand{\arraystretch}{1.5}
\begin{array}{l@{,\;\;}l}
a_0(\alpha) := \frac{1}{2} (\phantom{{-}}1 {+} e^{i \alpha}) \cos{\alpha} &
a_1(\alpha) := \frac{1}{2} (\phantom{{-}}1 {-} i e^{i \alpha})\sin{\alpha}, \\
a_2(\alpha) := \frac{1}{2} ({-}1 {+} e^{i \alpha}) \cos{\alpha} &
a_3(\alpha) := \frac{1}{2} ({-}1 {-} i e^{i \alpha}) \sin{\alpha}. \\
\end{array}
\renewcommand{\arraystretch}{1}
\]
Note that like in the previous example of the discrete Hartley
transforms in Section \ref{hartley}, we have used the property that
$F_N$ generates a finite group of order four to obtain the linear
combination shown in eq.~(\ref{fractFact}). 

It was shown in \cite{BM:96} that the one-parameter family
$\{F_{N, \alpha} \;|\; \alpha \in \R\} \subset {\C}^{N\times N}$ has the
following properties:
 
\begin{itemize}
\item[(i)] $F_{N, \alpha}$ is a unitary matrix for $\alpha \in \R$.
\item[(ii)] $F_{N, 0} = \onemat_N$ and $\DFT_{N, \pi/2} = \DFT_N$. 
\item[(iii)] $(F_{N,\alpha})^{1/\alpha} = \DFT_N$ for $\alpha \in
\R$. 
\item[(iv)] $F_{N, \alpha} \cdot \DFT_{N,\beta} = 
F_{N, {\alpha + \beta}}$, for
$\alpha, \beta \in R$. 
\end{itemize}

Using Theorem \ref{generic} we immediately obtain that $F_{2^n,
{\alpha}}$ can be computed in $O(n^2)$ elementary quantum operations
for all $\alpha \in \R$, since the complexity of the discrete Fourier
transform $F_{2^n}$ of length $2^n$ is $O(n^2)$
(cf. \cite{Coppersmith:94,Shor:94}) and the circulant matrix
appearing in this case is $C_\alpha :=
\zirc(a_0(\alpha),\ldots,a_3(\alpha))$ which can be implemented in
$O(1)$. More precisely we have
\[ C_\alpha = \zirc(\ket{\alpha}) = \DFT_4^{-1} \cdot
\diag(1, e^{-i \alpha}, e^{2 i \alpha}, e^{-i \alpha})
\cdot \DFT_4.
\]
Using the results of \cite{CW:2000} we can reduce the computational
complexity of $F_{2^n}$ to $O(n (\log n)^2 \log \log n)$ if we have no
restrictions on the number of ancilla qubits.

% Schaltkreis zur gebrochenen Fouriertransformation
% -------------------------------------------------
\begin{figure*}
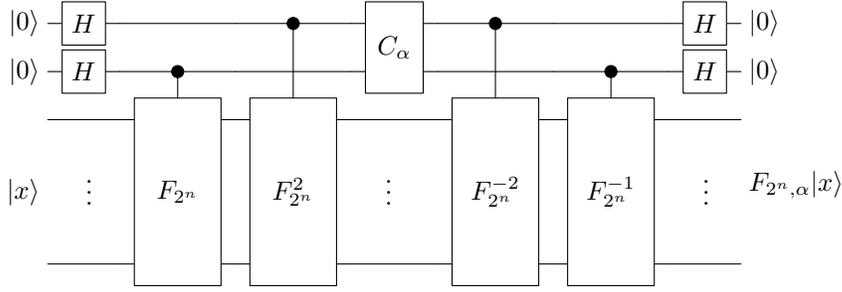

%\input{pic/fract.pic}
%\bigskip
\begin{center}
\begin{emp}(50,50)

  setunit 1.6mm; qubits(6); 
  ysave := QCycoord[2]-1/2QCheight; 
  label.lft(btex $|0\rangle$ etex,(QCxcoord,QCycoord[5])); 
  label.lft(btex $|0\rangle$ etex,(QCxcoord,QCycoord[4])); 
  label.lft(btex $|x\rangle$ etex,(QCxcoord,ysave)); 
  dropwire(1,2); ysave := ysave + 1mm;
  label(btex $\vdots$ etex,(QCxcoord+5.5mm,ysave));

  gate(gpos 2,btex $H$ etex, 3, btex $H$ etex);
  circuit(1.8*QCheight)(icnd 2,gpos 0,1,btex $F_{2^n}$ etex);
  circuit(1.8*QCheight)(icnd 3,gpos 0,1,btex $F_{2^n}^2$ etex);

  label(btex $\vdots$ etex,(QCxcoord+5.0mm,ysave));
  circuit(1.2*QCheight)(gpos 2,3,btex $C_\alpha$ etex);

  circuit(1.8*QCheight)(icnd 3,gpos 0,1,btex $F_{2^n}^{-2}$ etex);
  circuit(1.8*QCheight)(icnd 2,gpos 0,1,btex $F_{2^n}^{-1}$ etex);
  label(btex $\vdots$ etex,(QCxcoord+5.0mm,ysave));
  gate(gpos 2,btex $H$ etex, 3, btex $H$ etex);

  label.rt(btex $|0\rangle$ etex,(QCxcoord,QCycoord[3])); 
  label.rt(btex $|0\rangle$ etex,(QCxcoord,QCycoord[2])); 
  label.rt(btex $F_{2^n,\alpha}|x\rangle$ etex,(QCxcoord,ysave)); 
\end{emp}
\end{center}
\caption{\label{fract} 
Quantum circuit realizing a fractional quantum Fourier transform
}
\end{figure*}

Hence, we obtain the following theorem which summarizes the complexity
of computing a fraction Fourier transform.

\begin{theorem}
Let $\alpha\in \R$ and $F_{2^n, \alpha}$ be the fractional Fourier
transform of length $2^n$ and parameter $\alpha$. Then $\kappa_{\rm
anc}(F_{2^n, \alpha}) = O(n (\log n)^2 \log \log n)$. 
\end{theorem}

%%%%%%%%%%%%%%%%%%%%%%%%%%%%%%%%%%%%%%%%%%%%%%%%%%%%%%%%%%%%
%
% Subsection: Teleportation Revisited
%
%%%%%%%%%%%%%%%%%%%%%%%%%%%%%%%%%%%%%%%%%%%%%%%%%%%%%%%%%%%%

\goodbreak
\subsection{Teleportation Revisited}
\label{teleport}

In this section, we show how the well-known quantum circuit for
teleportation of an unknown quantum state (cf.~\cite{BBC:98,NC:2000})
can be interpreted with the help of our method. The essential feature
of the circuit in Figure~\ref{generic} is that a measurement on the upper
quantum bits can be carried out immediately after the transformation
$C_A$ has been performed.  This has the advantage that the
transformations $D(t_n)^{-1}$ etc. can be classically conditioned. This
explains the classical communication part of the teleportation
circuit.  To obtain the EPR states and the Bell measurement we use
some easy reformulations of the transformations exploiting the nature
of the projective circulant in case of $2\times 2$ matrices.

Suppose that Alice wants to teleport a quantum state
$\ket{\psi} = \alpha \ket{0} + \beta \ket{1}$ of a qubit in her
possession to a qubit in Bob's possession at a remote destination.  If
the destination qubit is in the state $\ket{0}$, then, conceptually,
the task is to apply a unitary operation $U$ such that
$U\ket{0}=\ket{\psi}$.  Specifically, the matrix $U$ can be chosen to
be of the form
\[ 
U = \left( \begin{array}{rr} \alpha & \overline{\beta} \\
\beta & -\overline{\alpha}\end{array} \right), \quad
\text{where} \; |\alpha|^2 + |\beta|^2 = 1.
\]
Clearly, it would not be feasible for Alice to communicate the
specification of $U$ to Bob by classical communication. Therefore, she
has to proceed in a different way.

Recall that the matrices
\[ 
\onemat_2 = \left(\begin{array}{cc} 1 & 0 \\ 0 & 1 \end{array} \right),\;
\sigma_x = \left(\begin{array}{cc} 0 & 1 \\ 1 & 0 \end{array} \right),\;
\sigma_z = \left(\begin{array}{rr} 1 & 0 \\ 0 & -1 \end{array} \right),\;
\sigma_x\sigma_z= \left(\begin{array}{ll} 0 & -1 \\ 1 & \phantom{-}0 
\end{array} \right).
\]
form a basis ${\cal
B}=\{\onemat_2,\sigma_x,\sigma_z,\sigma_x\sigma_z\}$ of the vector
space of complex $2\times 2$ matrices. Thus, the matrix $U$ can be
written as a linear combination $U=\sum_{B\in{\cal B}} c_B B$. 
Note that the Pauli basis ${\cal B}$ is an orthonormal
basis of $\C^{2\times 2}$ with respect to the inner product 
$\langle B | A \rangle =\frac{1}{2}\mbox{tr}(B^\dagger A)$. 
As a result, the coefficient
$c_B$ corresponding to $B\in {\cal B}$ can be easily computed by 
$c_B = \frac{1}{2} {\rm tr}(B^\dagger U)$. Consequently, we obtain
\begin{equation}\label{addTele}
U =  
\left(\frac{\alpha-\overline{\alpha}}{2} \right) \onemat_2 +
\left(\frac{\beta+\overline{\beta}}{2} \right) \sigma_x +
\left(\frac{\alpha+\overline{\alpha}}{2} \right) \sigma_z +
\left(\frac{\beta-\overline{\beta}}{2} \right) \sigma_x \sigma_z.
\end{equation}
The Pauli matrices define a projective representation of the abelian
group $\Z/2\Z \times \Z/2\Z$ (see, e.\,g., \cite{KR:2002a}). Applying the
method described in Section \ref{projCirculants}, the decomposition
(\ref{addTele}) gives rise to the projective circulant matrix $C_U$
defined as follows:
\[ 
C_U = \frac{1}{2} 
\left(
\begin{array}{rrrr}
\alpha - \overline{\alpha} & \beta + \overline{\beta} &
\alpha + \overline{\alpha} & \beta - \overline{\beta} \\[0.5ex]
\beta + \overline{\beta} & \alpha - \overline{\alpha} &
\beta - \overline{\beta} & \alpha + \overline{\alpha} \\[0.5ex]
\alpha + \overline{\alpha} & -(\beta - \overline{\beta}) &
\alpha - \overline{\alpha} & -(\beta + \overline{\beta}) \\[0.5ex]
\beta - \overline{\beta} & -(\alpha + \overline{\alpha}) &
\beta + \overline{\beta} & -(\alpha - \overline{\alpha}) 
\end{array}
\right).
\]
We can express $C_U$ by a sequence of  
Hadamard gates, controlled-not gates, and the
single-qubit gate $U$.  Indeed, a straightforward calculation 
shows that
\[ 
(H \otimes \onemat_2) \, C_U \, (H \otimes \onemat_2) =
\left(
\begin{array}{rrrr}
\alpha & 0 & 0 & \overline{\beta} \\[0.5ex]
\beta & 0 & 0 & -\overline{\alpha} \\[0.5ex]
0 & \beta & -\overline{\alpha} & 0 \\[0.5ex]
0 & \alpha & \overline{\beta} & 0 
\end{array} 
\right)
=: \widetilde{C_U}.
\]
Applying suitable permutations from the left and the right to the
matrix $\widetilde{C_U}$ we finally obtain the expression
\begin{equation}\label{cU}
\CNOT^{(1,2)}\,\CNOT^{(2,1)}\,\widetilde{C_U}\,\CNOT^{(2,1)} = U \otimes
\onemat_2.
\end{equation}
Overall we obtain that $C_U$ is given by the circuit shown in Figure
\ref{CUgate}. 

% Schaltkreis fuer C_U
% --------------------
\begin{figure}[hbt]
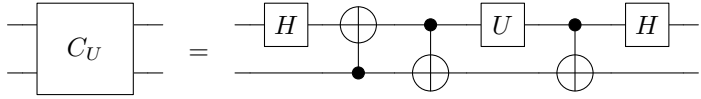

\begin{center}
\begin{emp}(50,50)
 setunit 1.6mm;
 qubits(2); 
 wires(2mm);
 circuit(2QCheight)(gpos 0,1, btex $C_U$ etex);
 wires(2mm);

 label(btex $=$ etex,(QCxcoord+1/2QCstepsize, QCycoord[0]+2.5mm));
 QCxcoord := QCxcoord + QCstepsize;

 wires(2mm);
 gate(gpos 1,btex $H$ etex);
 cnot(icnd 0, gpos 1);
 cnot(icnd 1, gpos 0);
 gate(gpos 1,btex $U$ etex);
 cnot(icnd 1, gpos 0);
 gate(gpos 1,btex $H$ etex);
 wires(2mm);
\end{emp}
\end{center}
\bigskip
\caption{\label{CUgate}
Realisation of the projective circulant $C_U$ in terms of the
operations $U$, controlled-not gates, and the Hadamard transform $H$.}
\end{figure}

We now turn to the circuit implementing the transformation $U$ using
the linear combination (\ref{addTele}). In the following we will 
modify the generic circuit step by step using elementary
identities of quantum gates. We start with the identity 

\bigskip

% Additiver Schaltkreis zur Teleportation
% ---------------------------------------

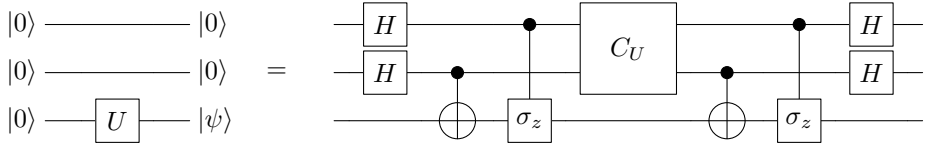
\begin{center}
\begin{emp}(50,50)
  setunit 1.6mm;
  qubits(3);
  label.lft(btex $|0\rangle$ etex,(QCxcoord,QCycoord[2]));
  label.lft(btex $|0\rangle$ etex,(QCxcoord,QCycoord[1]));
  label.lft(btex $|0\rangle$ etex,(QCxcoord,QCycoord[0]));
  wires(3*unit);
  gate(gpos 0, btex $U$ etex);
  wires(3unit);
  label.rt(btex $|0\rangle$ etex,(QCxcoord,QCycoord[2]));
  label.rt(btex $|0\rangle$ etex,(QCxcoord,QCycoord[1]));
  label.rt(btex $|\psi\rangle$ etex,(QCxcoord,QCycoord[0]));

  QCxcoord := QCxcoord + QCstepsize;
  label(btex $=$ etex,(QCxcoord+2mm, QCycoord[1]));
  QCxcoord := QCxcoord + QCstepsize;

  wires(2mm);
  gate(gpos 1, btex $H$ etex, 2, btex $H$ etex);
  cnot(icnd 1, gpos 0);
  gate(icnd 2, gpos 0, btex $\sigma_z$ etex);
  circuit(2QCheight)(gpos 1,2, btex $C_U$ etex);
  cnot(icnd 1, gpos 0);
  gate(icnd 2, gpos 0, btex $\sigma_z$ etex);
  gate(gpos 1, btex $H$ etex, 2, btex $H$ etex);
  wires(2mm);
\end{emp} 
\end{center}

\medskip

\noindent which is obtained directly from the method of Sections
\ref{design} and \ref{projCirculants}.  The matrix $U$ is given by a linear combination of Pauli matrices in 
eq.~(\ref{addTele}). This limear combination determines $C_U$. 
We rewrite $C_U$  as a
product of CNOT gates and local unitary transformations using eq.~(\ref{cU}). 
We obtain the circuit 

\bigskip

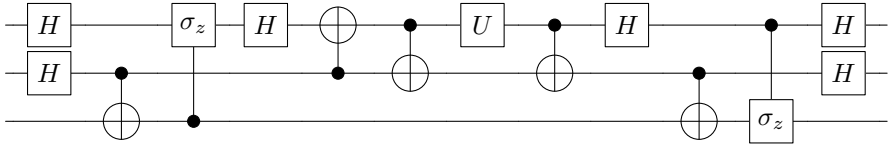
\begin{center}
\begin{emp}(50,50)
  setunit 1.6mm;
  qubits(3);
%  label(btex $=$ etex,(QCxcoord+1/2QCstepsize, QCycoord[1]));
  QCxcoord := QCxcoord + QCstepsize;

  wires(1mm);
  gate(gpos 1, btex $H$ etex, 2, btex $H$ etex);
  cnot(icnd 1, gpos 0);
  gate(icnd 0, gpos 2, btex $\sigma_z$ etex);
  gate(gpos 2, btex $H$ etex);
  cnot(icnd 1, gpos 2);
  cnot(icnd 2, gpos 1);
  gate(gpos 2, btex $U$ etex);
  cnot(icnd 2, gpos 1);
  gate(gpos 2, btex $H$ etex);
  cnot(icnd 1, gpos 0);
  gate(icnd 2, gpos 0, btex $\sigma_z$ etex);
  gate(gpos 1, btex $H$ etex, 2, btex $H$ etex);
  wires(1mm);
\end{emp} 
\end{center}

\medskip

\noindent where we also turned the first controlled-$\sigma_z$ gate 
upside down. Now we can use the basic fact that $H
\sigma_z H = \sigma_x$ to rewrite the controlled-$\sigma_z$ framed by
the two Hadamard gates on the top wire in the following way:

\bigskip

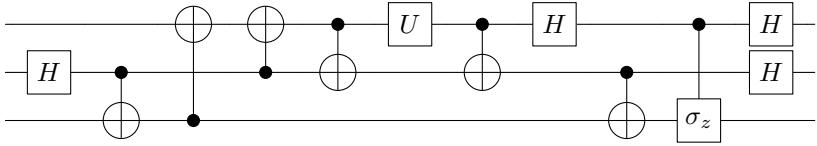
\begin{center}
\begin{emp}(50,50)
  setunit 1.6mm;
  qubits(3);
%  label(btex $=$ etex,(QCxcoord+1/2QCstepsize, QCycoord[1]));
  QCxcoord := QCxcoord + QCstepsize;

  wires(1mm);
  gate(gpos 1, btex $H$ etex);
  cnot(icnd 1, gpos 0);
  cnot(icnd 0, gpos 2);
  cnot(icnd 1, gpos 2);
  cnot(icnd 2, gpos 1);
  gate(gpos 2, btex $U$ etex);
  cnot(icnd 2, gpos 1);
  gate(gpos 2, btex $H$ etex);
  cnot(icnd 1, gpos 0);
  gate(icnd 2, gpos 0, btex $\sigma_z$ etex);
  gate(gpos 1, btex $H$ etex, 2, btex $H$ etex);
  wires(1mm);
\end{emp} 
\end{center}

\medskip

We can simplify the sequence of the first five gates of
the last circuit. Indeed, since we start from the state
$\ket{0}\ket{0}\ket{0}$, the state resulting from applying the first
five gates is $\ket{0}(\ket{00} + \ket{11})$. This state can be
obtained by applying the first two of these five gates alone. 
This simplification yields the following circuit

\bigskip

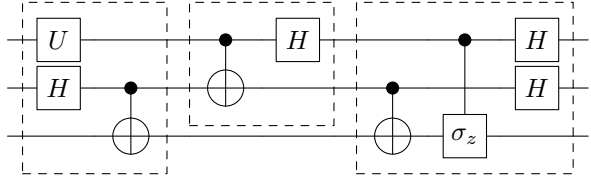
\begin{center}
\begin{emp}(50,50)
  setunit 1.6mm;
  qubits(3);
%  label(btex $=$ etex,(QCxcoord+1/2QCstepsize, QCycoord[1]));
  QCxcoord := QCxcoord + QCstepsize;

  wires(2mm);
  draw (QCxcoord, QCycoord[0]-5mm) -- (QCxcoord, QCycoord[2]+5mm) 
       -- (QCxcoord+2QCstepsize, QCycoord[2]+5mm) 
       -- (QCxcoord+2QCstepsize, QCycoord[0]-5mm)
       -- cycle dashed evenly;
  gate(gpos 1, btex $H$ etex, 2, btex $U$ etex);
  cnot(icnd 1, gpos 0);
  wires(3mm);
  draw (QCxcoord, QCycoord[1]-5mm) -- (QCxcoord, QCycoord[2]+5mm) 
       -- (QCxcoord+2QCstepsize, QCycoord[2]+5mm) 
       -- (QCxcoord+2QCstepsize, QCycoord[1]-5mm)
       -- cycle dashed evenly;
  cnot(icnd 2, gpos 1);
  gate(gpos 2, btex $H$ etex);
  wires(3mm);
  draw (QCxcoord, QCycoord[0]-5mm) -- (QCxcoord, QCycoord[2]+5mm) 
       -- (QCxcoord+3QCstepsize, QCycoord[2]+5mm) 
       -- (QCxcoord+3QCstepsize, QCycoord[0]-5mm)
       -- cycle dashed evenly;
  cnot(icnd 1, gpos 0);
  gate(icnd 2, gpos 0, btex $\sigma_z$ etex);
  gate(gpos 1, btex $H$ etex, 2, btex $H$ etex);
  wires(2mm);
\end{emp} 
\end{center}

\medskip

\noindent which decomposes into three stages: (i) {\em EPR pair and state
preparation} in which, starting from the ground state
$\ket{0}\ket{0}\ket{0}$, two of the bits are turned into an EPR state
while the third qubit holds the unknown state $\ket{\psi}$, (ii) {\em
Bell measurement} of the two most significant qubits, and (iii) a {\em
reconstruction} operation which is a conditional transformation on
qubit one depending on the outcome of the measurement of qubits two
and three.

\bigskip

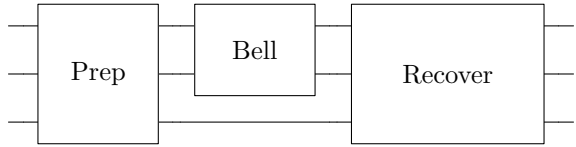
\begin{center}
\begin{emp}(50,50)
  setunit 1.6mm;
  qubits(3);
%  label(btex $=$ etex,(QCxcoord+1/2QCstepsize, QCycoord[1]));
  QCxcoord := QCxcoord + QCstepsize;

  wires(2mm);
  circuit(2.5QCheight)(gpos 0, 1,2, btex Prep etex);
  wires(1mm);
  circuit(2.5QCheight)(gpos 1,2, btex Bell etex);
  wires(1mm);
  circuit(4QCheight)(gpos 0, 1,2, btex Recover etex);
  wires(2mm);
\end{emp} 
\end{center}

\medskip

Hence, this circuit equals the teleportation circuit for an unknown
quantum state $\ket{\psi} = U \ket{0}$, see for instance 
\cite[Section~1.3.7]{NC:2000} or \cite{BBC:98}. In summary, we have
seen that it is possible to derive a transformation of a quantum
circuit corresponding to linear combination of the transformation~$U$
as a sum of Pauli matrices into the teleportation circuit.

%%%%%%%%%%%%%%%%%%%%%%%%%%%%%%%%%%%%%%%%%%%%%%%%%%%%%%%%%%%%
%
% Section: Conclusions and Outlook
%
%%%%%%%%%%%%%%%%%%%%%%%%%%%%%%%%%%%%%%%%%%%%%%%%%%%%%%%%%%%%

\section{Conclusions}
The factorization of a unitary matrix in terms of elementary quantum
gates amounts to solve a word problem in a unitary group. This problem
is quite difficult, in particular since only words of small length,
which correspond to efficient algorithms, are of practical interest.
Few methods are known to date for the design of quantum
circuits. Several ad-hoc methods for quantum circuit design have been
proposed, mostly heuristic search techniques based on genetic
algorithms, simulated annealing or the like.  Such methods are
confined to fairly small circuit sizes, and the solutions produced by
such heuristics are typically difficult to interpret.

The method presented in this paper follows a completely different
approach. We assume that we have a set of efficient
quantum circuits available. Our philosophy is to reuse and combine
these circuits to build a new quantum circuit. We have developed a
sound mathematical theory, which allows to solve such problems under
certain well-defined conditions.  Following this approach, we have
demonstrated that the discrete Hartley transforms and fractional
Fourier transforms have extremely efficient realizations on a quantum
computer.  

It should be stressed that the method is by no means exhausted by
these examples. From a practical point of view, it would be
interesting to build a database of moderately sized matrix groups
which have efficient quantum circuits. This database could in turn be
searched for a given transformation by means of linear algebra.  It is
an appealing possibility to automatically derive quantum circuit
implementations in this fashion.

\nix{
It seems possible to apply the method presented to transformations
used as observables in quantum algorithms for which by now there have
been no efficient factorizations being found. An interesting question
is whether the method can be used as to derive approximate
factorizations of matrices.

It remains to find further classes of signal transformations which
become efficiently implementable by the presented method and to build
a database of moderately sized matrix groups which can have efficient
quantum circuits. .
}
 
\nonumsection{Acknowledgements} The research of A.K. has been partly
supported by NSF grant EIA 0218582, and by a Texas A\&M TITF
grant. Part of this work has been done while M.R.\ was at the
Institute for Algorithms and Cognitive Systems, University of
Karlsruhe, Karlsruhe, Germany, and during a visit to the Mathematical
Sciences Research Institute, Berkeley, USA.
He wishes to thank both institutions for their hospitality. 
His research has been supported by the European Community
under contract IST-1999-10596 (Q-ACTA), CSE, and MITACS.

\nonumsection{References}

%References are to be listed in the order cited in the text. Use the style
%shown in the following examples. For journal names, use the standard
%abbreviations. Typeset references in 9 pt Times Roman.

\end{empfile}

%%%%%%%%%%%%%%%%%%%%%%%%%%%%%%%%%%%%%%%%%%%%%%%%%%%%%%%%%%%%
% 
% The literature
%
%%%%%%%%%%%%%%%%%%%%%%%%%%%%%%%%%%%%%%%%%%%%%%%%%%%%%%%%%%%%
 
%\nocite{*}

%\begin{thebibliography}{9}
%
%\bibitem{B} M. J. Beeson, {\it Foundations of Constructive Mathematics}
%(Springer, Berlin, 1985).
%\bibitem{C} K. L. Clark, ``Negations as failure,'' in {\it Logic and Data
%Bases}, eds. H. Gallaire and J. Winker (Plenum Press, New York, 1973) pp.
%293--306.
%\end{thebibliography}


\begin{thebibliography}{10}

\bibitem{BBC+:95}
A.~Barenco, C.~H. Bennett, R.~Cleve, D.~P. DiVincenzo, N.~Margolus, P.~Shor,
  T.~Sleator, J.~A. Smolin, and H.~Weinfurter.
\newblock {Elementary gates for quantum computation}.
\newblock {\em Physical Review~A}, 52(5):3457--3467, November 1995.

\bibitem{ABH+:2001}
G.~Alber, Th. Beth, M.~Horodecki, P.~Horodecki, R.~Horodecki, M.~R{\"o}tteler,
  H.~Weinfurter, R.~Werner, and A.~Zeilinger.
\newblock {\em Quantum Information: An Introduction to Basic Theoretical
  Concepts and Experiments}, volume 173 of {\em Springer Texts in Modern
  Physics}.
\newblock Springer, 2001.

\bibitem{Shor:94}
P.~W. Shor.
\newblock {Algorithms for quantum computation: discrete logarithm and
  factoring}.
\newblock In {\em Proc. FOCS 94}, pages 124--134. IEEE Computer Society Press,
  1994.

\bibitem{CW:2000}
R.~Cleve and J.~Watrous.
\newblock {Fast parallel circuits for the quantum Fourier transform}.
\newblock Technical report, LANL preprint quant-ph/0006004, 2000.

\bibitem{Ettinger:00}
M.~Ettinger.
\newblock {Quantum time-frequency transforms}.
\newblock LANL preprint quant--ph/0005134, 2000.

\bibitem{Hoyer:97}
P.~H{\o}yer.
\newblock {Efficient quantum transforms}.
\newblock LANL preprint quant--ph/9702028, February 1997.

\bibitem{Klappenecker:99}
A.~Klappenecker.
\newblock Wavelets and wavelet packets on quantum computers.
\newblock In M.A. Unser, A.~Aldroubi, and A.F. Laine, editors, {\em Wavelet
  Applications in Signal and Image Processing {VII}}, pages 703--713. SPIE,
  1999.

\bibitem{KR:2001}
A.~Klappenecker and M.~R{\"o}tteler.
\newblock {Discrete cosine transforms on quantum computers}.
\newblock In {\em Proc. IEEE R8-EURASIP Symposium on Image and Signal
  Processing and Analysis (ISPA01)}, Pula, Croatia, 2001.

\bibitem{PRB:99}
M.~P{\"u}schel, M.~R{\"o}tteler, and Th. Beth.
\newblock {Fast quantum Fourier transforms for a class of non-abelian groups}.
\newblock In {\em Proceedings Applied Algebra, Algebraic Algorithms and
  Error-Correcting Codes (AAECC-13)}, volume 1719 of {\em Lecture Notes in
  Computer Science}, pages 148--159. Springer, 1999.

\bibitem{beth:89}
Th. Beth.
\newblock {Generating fast Hartley transforms - another application of the
  algebraic discrete Fourier transform}.
\newblock In {\em Proc. URSI-ISSSE '89}, pages 688--692, 1989.

\bibitem{Bracewell:79}
Bracewell.
\newblock {\em The Hartley Transform}.
\newblock Cambridge Univ. Press, 1979.

\bibitem{CB:93}
M.~Clausen and U.~Baum.
\newblock {\em Fast Fourier Transforms}.
\newblock BI-Verlag, 1993.

\bibitem{Davis:79}
P.~Davis.
\newblock {\em Circulant matrices}.
\newblock Wiley-Interscience New York, 1979.

\bibitem{Huppert:79}
B.~Huppert.
\newblock {\em Endliche Gruppen I}.
\newblock Grundlehren der mathematischen Wissenschaften. Springer-Verlag,
  Berlin, 1979.

\bibitem{DiVincenzo:98}
D.~DiVincenzo.
\newblock Quantum gates and circuits.
\newblock {\em Proc. R. Soc. London A}, 454(1969):261--276, 1998.

\bibitem{serre96}
J.P. Serre.
\newblock {\em Linear Representations of Finite Groups}, volume~42 of {\em
  GTM}.
\newblock Springer-Verlag, Berlin, 1996.
\newblock (5th corr.\ printing).

\bibitem{KR:2002a}
A.~Klappenecker and M.~R{\"o}tteler.
\newblock {Beyond Stabilizer Codes I: Nice Error Bases}.
\newblock {\em IEEE Transactions on Information Theory}, 48(8):2392--2395,
  2002.

\bibitem{Schur:09}
I.~Schur.
\newblock {Beitr{\"a}ge zur Theorie der Gruppen linearer homogener
  Substitutionen}.
\newblock {\em Trans. Amer. Math. Soc.}, 10:159--175, 1909.

\bibitem{Kitaev:95}
A.~Yu. Kitaev.
\newblock {Quantum measurements and the abelian stabilizer problem}.
\newblock LANL preprint quant--ph/9511026, November 1995.

\bibitem{Kitaev:97}
A.~Yu. Kitaev.
\newblock Quantum computations: algorithms and error correction.
\newblock {\em Russian Math. Surveys}, 52(6):1191--1249, 1997.

\bibitem{CEMM:98}
{Cleve, R. and Ekert, A. and Macchiavllo, C. and Mosca, M.}
\newblock Quantum algorithms revisited.
\newblock {\em Proceedings of the Royal Society of London (Series A)},
  454(1969):339--354, 1998.

\bibitem{Jozsa:98}
R.~Jozsa.
\newblock {Quantum algorithms and the Fourier transform}.
\newblock {\em Proc. R. Soc. Lond. A}, 454:323--337, 1998.

\bibitem{Grigoriev:96}
D.~Grigoriev.
\newblock {Testing shift-equivalence of polynomials using quantum machines}.
\newblock In {\em {Proc. International Symposium on Symbolic and Algebraic
  Computation (ISSAC)}}, pages 49--54. ACM Press, 1996.

\bibitem{BM:96}
S.~Balasubramaiam and J.~McClellan.
\newblock {The discrete rotational Fourier transform}.
\newblock {\em IEEE Transactions on Signal Processing}, 44(4):994--998, 1996.

\bibitem{CKO:2000}
C.~Candan, M.~Kutay, and H.~Ozaktas.
\newblock {The discrete fractional Fourier transform}.
\newblock {\em IEEE Transactions on Signal Processing}, 48(5):1329--1337, 2000.

\bibitem{Coppersmith:94}
D.~Coppersmith.
\newblock {An approximate Fourier transform useful for quantum factoring}.
\newblock Technical Report RC~19642, IBM Research Division, 1994.
\newblock see also LANL preprint quant-ph/0201067.

\bibitem{BBC:98}
G.~Brassard, S.~Braunstein, and R.~Cleve.
\newblock {Teleportation as a quantum computation}.
\newblock {\em {Physica D}}, 120(1--2):43--47, 1998.

\bibitem{NC:2000}
M.~Nielsen and I.~Chuang.
\newblock {\em Quantum Computation and Quantum Information}.
\newblock Cambridge University Press, 2000.

\end{thebibliography}
\end{document}